\documentclass[a4paper]{jpconf}
\usepackage{graphicx}

\usepackage{iopams}

\begin{document}
\title{Neutral bions in the ${\mathbb C}P^{N-1}$ model for 
resurgence}

\author{Tatsuhiro Misumi$^{1}$, Muneto Nitta$^{2}$, and 
Norisuke Sakai$^{3}$}

\address{Department of Physics, and Research and Education 
Center for Natural Sciences, 
Keio University, Hiyoshi 4-1-1, Yokohama, Kanagawa 223-8521, Japan
}

\ead{$^{1}$misumi@phys-h.keio.ac.jp, $^{2}$nitta@phys-h.keio.ac.jp, 
$^{3}$norisuke.sakai@gmail.com (Speaker)
}

\begin{abstract}
Classical configurations in the ${\mathbb C}P^{N-1}$ model 
on ${\mathbb R}^{1}\times S^{1}$ is studied with twisted 
boundary conditions. 
Starting from fractional instantons with 
the ${\mathbb Z}_{N}$ twisted boundary conditions, 
we review briefly the relevance of our study to resurgence 
phenomenon in field theory. 
We consider primarily configurations composed of multiple 
fractional instantons, called ``neutral bions", which are 
identified as ``perturbative infrared renormalons". 
We construct an explicit ansatz corresponding to topologically 
trivial configurations containing one fractional instanton 
($\nu=1/N$) and one fractional anti-instanton ($\nu=-1/N$), 
which is guaranteed to become a solution of field 
equations asymptotically for large separations. 
The attractive interactions between the instanton 
constituents from small to large separations are found 
to be consistent with the standard 
separated-instanton calculus. 
Our results suggest that the ansatz enables us to study bions 
and the related physics for a wide range of separations. 
This talk is mainly based on our recent work published in 
JHEP {\bf 1406}, 164 (2014)  [arXiv:1404.7225 [hep-th]].
\end{abstract}

\section{Introduction}

Recently a progress has been brought about to understand the 
confinement mechanism in QCD-like theories by compactifying 
the theories in one spatial dimension with the period $L$.  
Because of the asymptotic freedom, the small $L$ theory is in the 
weak-coupling regime ($L\ll1/\Lambda_{\rm QCD}$) in QCD(adj.) 
on ${\mathbb R}^{3} \times S^{1}$. 
Instanton is a particle-like soliton of codimension $D$ in 
$D$-dimensional spacetime. 
If ${\mathbb Z}_{N}$ twisted boundary conditions are imposed, 
we can obtain fractional instantons which have fractional 
amount of instanton number. 
A molecule of fractional instantons and anti-instantons are 
generally called bions. 
For compactified space with small $L$, the perturbative 
analysis is reliable and shows that the Wilson holonomy stabilizes 
the ${\mathbb Z}_{N}$ twisted boundary condition. 
In this case, one can demonstrate the condensation of classical 
field configurations composed of fractional instantons and 
anti-instantons with magnetic charge, which are called magnetic 
bions. 
Condensation of magnetic bions implies the confinement in 
the compactified theory \cite{Unsal:2007vu, Unsal:2007jx, Shifman:2008ja, 
Poppitz:2009uq, Anber:2011de,
Poppitz:2012sw}. 
It is anticipated that this phenomenon continues to hold 
for large $L$, suggesting the confinement even for uncompactified 
theory.

It has also been pointed out that neutral bions play an 
important role in explaining the resurgence phenomenon 
\cite{Argyres:2012vv, Dunne:2012ae, Dabrowski:2013kba, Dunne:2013ada, 
Cherman:2013yfa, Basar:2013eka, Dunne:2014bca, Cherman:2014ofa}. 
Neutral bions are composite of fractional instantons and 
anti-instantons with vanishing topological charge (instanton number) 
and zero magnetic charge. 
Although perturbation series in field theories are divergent, 
the Borel transformation is useful to obtain the re-summation 
of such divergent series. 
The Borel transform often exhibits singularities on the 
positive real axis producing an imaginary ambiguities in 
the Borel-resummed results. 
It has been found that these ambiguities in quantum mechanics 
are cancelled by contributions from molecule of 
instanton and anti-instantons \cite{Bogomolny:1980ur, 
ZinnJustin:1981dx, ZinnJustin:2004ib,ZinnJustin:1980uk}. 
However, asymptotically free field theories like QCD and the 
${\mathbb C}P^{N-1}$ nonlinear sigma models exhibits similar 
but more serious singularities near the origin, which are 
called infrared renormalons \cite{'tHooft:1977am,Beneke:1998ui}. 
Recently it has been found that the nonperturbative 
contributions around the neutral saddle points also have 
imaginary ambiguities. 
These imaginary ambiguities of bion amplitudes are found to 
cancel \cite{Argyres:2012vv,Dunne:2012ae} precisely the imaginary 
ambiguities associated with the infrared renormalons, 
resulting in a more rigorous foundation of field theory. 
This phenomenon is called the resurgence. 
It is expected that full semi-classical expansion including 
perturbative and non-perturbative sectors, which is called 
resurgent expansion \cite{Ec1}, leads to unambiguous and 
self-consistent definition of field theories in the spirit of 
quantum mechanical examples \cite{Bogomolny:1980ur, 
ZinnJustin:1981dx, ZinnJustin:2004ib}. 
However, it is not straightforward to verify these arguments 
in gauge theories directly, since it is difficult to find an 
explicit ansatz of bion configurations.

At this stage, it is quite interesting and useful to study 
simpler models in lower dimensions 
\cite{Dunne:2012ae, Dabrowski:2013kba} 
instead of gauge theories in four spacetime dimensions. 
In particular, models in two spacetime dimensions are more tractable 
compared to gauge theories in four dimensions, but have many 
features which are common to gauge theories in four dimensions. 
Among them, the ${\mathbb C}P^{N-1}$ model in two spacetime 
dimensions has been studied as a toy model \cite{Polyakov} of 
the Yang-Mills theory in four spacetime dimensions, because of 
their interesting features such as the dynamical mass gap, 
the asymptotic freedom and the existence of instantons as 
Bogomol'nyi-Prasad-Sommerfield (BPS) solutions \cite{Polyakov:1975yp}.
The ${\mathbb C}P^{N-1}$ model on ${\mathbb R}^1 \times S^1$ 
with twisted boundary conditions admits BPS fractional instantons 
as configurations with the minimal topological charge 
\cite{Eto:2004rz,Eto:2006mz,Eto:2007aw,Eto:2006pg} (see also 
Refs.~\cite{Bruckmann:2007zh,Brendel:2009mp,Harland:2009mf}). 
In Ref.~\cite{Dunne:2012ae}, generic arguments on bion 
configurations were given in the ${\mathbb C}P^{N-1}$ model 
on ${\mathbb R}^1 \times S^1$ with ${\mathbb Z}_{N}$ twisted 
boundary conditions, based on the independent instanton 
description taking account of interactions between 
far-separated fractional instantons and anti-instantons. 
More recently, exact non-BPS solutions were found 
in the ${\mathbb C}P^{N-1}$ model on ${\mathbb R}^1 \times S^1$
with the ${\mathbb Z}_{N}$ twisted boundary condition 
\cite{Dabrowski:2013kba}. 
The simplest non-BPS solution that they found is a four-instanton 
configuration composed of two fractional instantons 
(instanton charge $\nu=1/N$) and two fractional anti-instantons 
($\nu=-1/N$) for $N\geq3$ placed at particular relative positions. 
Their study suggests that it is likely that bion configurations needed 
to understand the resurgence phenomenon may not be solutions of field 
equations. 
It has been known that field configurations other than the solution 
of the equations of motion may give significant contributions 
in the functional integral. 
Recently we constructed Ansatz of bion configurations that are 
guaranteed to become solutions of the field equations 
asymptotically as separations between constituent fractional 
instantons get larger, and studied interactions of constituent 
fractional instantons in bion configuration \cite{Misumi:2014jua}. 
More recently, we have also studied bions in the Grassmann 
sigma models which include the ${\mathbb C}P^{N-1}$ model 
as a subclass \cite{Misumi:2014bsa}. 

\section{${\mathbb C}P^{N-1}$ model}
\label{sec:CP}

Since bions in nonlinear sigma models in two spacetime dimensions 
are interesting and more tractable, let us consider the 
${\mathbb C}P^{N-1}$ model.  
To describe the target space 
${\mathbb C}P^{N-1}=\frac{SU(N)}{SU(N-1)\times U(1)}$, 
we denote an $N$-component vector of 
complex scalar fields as $\omega(x)$, and 
a normalized complex $N$-component vector 
composed from $\omega$ as $n(x)$ 
\begin{equation}
n(x) \equiv \omega(x)/|\omega(x)|, \quad 
|\omega|=\sqrt{\omega^\dagger \omega}. 
\end{equation}
The Lagrangian in Euclidean two dimensions for the nonlinear 
sigma model with ${\mathbb C}P^{N-1}$ as the target space 
is given by 
\begin{equation}
{\cal L}={1\over{2\pi g^{2}}} (D_{\mu}n)^{\dagger} (D_{\mu} n)\,,
\end{equation}
and the topological charge representing 
$\pi_2({\mathbb C}P^{N-1}) \simeq {\mathbb Z}$ 
is defined by 
\begin{equation}
Q=\frac{1}{2\pi}\int d^{2}x \; i\epsilon_{\mu\nu} 
(D_{\mu}n)^{\dagger} (D_{\nu} n)
=\frac{1}{2\pi}\int d^{2}x \epsilon_{\mu\nu}\partial_{\mu} A_{\nu}\,,
\label{Qdef}
\end{equation}
where $d^{2} x\equiv dx_{1}dx_{2}$ and $\mu,\nu=1,2$. 
Our convention of the covariant derivative is 
$D_{\mu}=\partial_{\mu}-iA_{\mu}$ with a composite gauge 
field 
\begin{equation}
A_{\mu} (x) \equiv -in^{\dag}\partial_{\mu}n. 
\label{eq:composite_gauge_field}
\end{equation}

We consider ${\mathbb R}^{1}\times S^{1}$ as the geometry of 
base manifold, and configurations on it satisfying periodicity 
in the $x_2$ direction with the period $L$. 
The Lagrangian ${\cal L}$ and topological charge $Q$ can be expressed 
in terms of the projection operator 
${\bf P}\equiv nn^{\dagger}= 
{\omega \omega^{\dagger}\over{\omega^{\dagger}\omega}}$ 
and using the complex coordinate $z \equiv x_1+ix_2$, 
\begin{eqnarray}
{\cal L}&=&{2\over{g^{2}}}
{\rm Tr} \left[ \partial_{z}{\bf P}\partial_{\bar z} {\bf P} \right]\,,\\
Q&=&2\int d^2 x {\rm Tr} \left[{\bf P} (\partial_{\bar z}{\bf P}
\partial_{z} {\bf P}-\partial_{z}{\bf P}\partial_{\bar z} {\bf P}) \right]\,.
\end{eqnarray}
We define the euclidean energy density (Lagrangian) $s(x_1)$, 
and the topological charge density $q(x_1)$ as functions of 
$x_{1}$ after the integration over $x_2$:
\begin{eqnarray}
s(x_{1})&=&{1\over{g^{2}\pi}}\int dx_{2} 
{\rm Tr} \left[ \partial_{z}{\bf P}\partial_{\bar z} {\bf P} \right]\,,
\\
q(x_{1})&=&{1\over{\pi}}\int dx_{2} {\rm Tr} \left[{\bf P} 
(\partial_{\bar z}{\bf P}\partial_{z} {\bf P}-\partial_{z}
{\bf P}\partial_{\bar z} {\bf P}) \right]\,.
\end{eqnarray}

We can perform the Bogomol'nyi completion to obtain the 
Bogomol'nyi bound \cite{Bogomolny:1975de} for the total energy $S$ 
\begin{equation}
S={1\over2\pi{g^{2}}}\int d^{2}x \left(\frac{1}{2}\left|
D_{\mu}n \mp i \epsilon_{\mu\nu}D_{\nu} n\right|^2
\pm i \epsilon_{\mu\nu} (D_{\nu}n)^{\dag} (D_{\mu} n))\right)
\ge
\pm \frac{
Q}{g^2}. 
\end{equation}
The Bogomol'nyi bound with the upper sign is saturated when 
the following BPS equation is satisfied  
\begin{equation}
D_{1}n - i D_{2} n=
\frac{1}{2}D_{\bar z}n=0.  
\label{eq:BPSeq}
\end{equation}
The solution of the BPS equation is precisely holomorphic 
$\omega(z)$. 
The Bogomol'nyi bound with the lower sign is saturated when 
the following anti-BPS equation is satisfied  
\begin{equation}
D_{1}n + i D_{2} n=
\frac{1}{2}D_{\bar z}n=0.  
\end{equation}
The solution of the anti-BPS equation is 
anti-holomorphic $\omega(\bar z)$. 

In the following we omit the coupling $1/g^2$ for simplicity 
unless stated otherwise.

\section{Borel resummation 
} 

Let us give a brief review of resurgence and the importance of bion 
contributions. 
It has been known that the number of Feynman diagrams in quantum 
field theory grows factorially. 
This factorially divergent perturbation series of quantum field 
theory is usually discussed by means of the Borel transform. 
The Borel transform method is applicable to the following 
class of divergent series (called Gevrey-1) 
\begin{equation}
  P(g^2) = \sum_{q=0}^\infty a_{q} (g^{2})^q, \quad 
 |a_{q}| \le C q! \left(\frac{1}{A}\right)^q, 
\end{equation}
where $C, A$ are constants. 
The Borel transform $BP(t)$ is defined as 
\begin{equation}
BP(t) = \sum_{q=0}^\infty \frac{a_{q}}{q!} t^q ,
\end{equation}
and the Borel resummation ${\mathbb B}(g^2)$ 
is defined as  
\begin{equation}
{\mathbb B}(g^2) = \int_0^\infty \frac{dt}{g^2} 
e^{-t/g^2} BP(t). 
\end{equation}
One can easily see that the Borel resummation ${\mathbb B}(g^2)$ 
reproduces the original sum $P(g^2)$ correctly whenever one can 
exchange the integral and the sum. 
Otherwise, we need to define the sum in terms of the Borel 
resummation.

As a simplified toy model, let us consider a factorially 
divergent series of the following one with alternating signs 
\begin{equation}
P(g^2)
= C \sum_{q=0}^\infty q! \left(\frac{-g^{2}}{A}\right)^q. 
\end{equation}
Then the Borel transform becomes an analytic function 
without singularities on the positive real axis 
\begin{equation}
BP(t)=C\sum_{q=0}^\infty \left(\frac{-t}{A}\right)^q
=\frac{CA}{A+t}. 
\end{equation}
Therefore the Borel resummation is well-defined as an integral 
along the positive real axis 
\begin{equation}
{\mathbb B}
(g^2) = \int_0^\infty \frac{dt}{g^2}e^{-t/g^2}
\frac{CA}{A+t}. 
\end{equation}
This alternating factorially divergent series is a typical example 
of Borel summable divergent series. 

On the other hand, if perturbation series is not alternating, 
the factorially divergent series gives the Borel transform with 
singularities on positive real axis and the Borel resummation 
has imaginary ambiguities. 
For instance, suppose that the perturbation series 
$P_{\rm pert}(g^2)$ gives 
non-alternating factorially divergent series like 
\begin{equation}
P_{\rm pert}(g^2)
= C \sum_{q=0}^\infty q! \left(\frac{g^{2}}{A}\right)^q. 
\end{equation}
The Borel transform has a singularity on positive real axis 
\begin{equation}
BP_{\rm pert}(t)=C\sum_{q=0}^\infty \left(\frac{t}{A}\right)^q
=\frac{CA}{A-t}, 
\end{equation}
\begin{equation}
{\mathbb B}_{\rm pert}(g^2) = \int_0^\infty \frac{dt}{g^2}e^{-t/g^2}
\frac{CA}{A-t}. 
\end{equation}
Therefore the Borel resummation has imaginary ambiguities depending on 
the choice of integration contours to avoid the singularity 
(equivalently the analytic continuation from negative real axis) 
\begin{equation}
{\mathbb B}_{\rm pert}(g^2\pm \epsilon) = 
{\rm Re}{\mathbb B}_{\rm pert}(|g^2|) 
\pm i {\rm Im}{\mathbb B}_{\rm pert}(|g^2|), 
\; \; 
{\rm Im}{\mathbb B}_{\rm pert}(|g^2|)\sim -\pi e^{-A/g^2}. 
\end{equation}
This ambiguous imaginary part 
has to be cancelled by contributions 
from other saddle points in the path-integral. 
In quantum mechanics or scalar field theories, the position of 
the singularity in the Borel plane has been found to correspond 
to molecules of instanton and anti-instanton 
\cite{Bogomolny:1980ur,ZinnJustin:1981dx,ZinnJustin:2004ib}. 
However, the asymptotically free theories like QCD and the 
${\mathbb C}P^{N-1}$ model give additional singularities 
due to IR renormalons which are much closer to the origin 
and hence give much larger contributions. 
Recent study \cite{Argyres:2012vv,Dunne:2012ae} 
showed that the bion amplitudes give 
the nonperturbative contribution needed to cancel the 
imaginary ambiguity due to the IR renormalons.

\section{Fractional instantons and neutral-bion in ${\mathbb Z}_{N}$ twisted boundary conditions}

When one space direction is compactified to $S^1$ with 
the circumference $L$, there is a possibility to impose 
non-periodic boundary condition. 
Moreover, asymptotic free theory becomes weakly coupled, 
and the stable configurations of Wilson holonomies of $U(1)^{N-1}$ 
subgroup of $SU(N)$ can be analyzed perturbatively 
\cite{Dunne:2012ae, Dabrowski:2013kba}. 
It turns out that the following $Z_N$ twisted boundary 
condition is most favorable under certain conditions 
\begin{equation}
\omega(x_{1}, x_{2}+L) = \Omega \,\omega(x_{1},x_{2})\,,
\,\,\,\,\,\,\,\,\,\,\,\,\,
\Omega={\rm diag.}\left[1, e^{2\pi i/N}, e^{4\pi i/N},\cdot\cdot\cdot, e^{2(N-1)\pi i/N}  \right]\,.
\label{ZNC}
\end{equation}
This ${\mathbb Z}_{N}$ twisted boundary condition 
corresponds to the vacuum with the symmetry breaking 
$SU(N) \to U(1)^{N-1}$. 
The Wilson-loop holonomy in the compactified direction is given by
\begin{equation}
\langle A_{2} \rangle = (0, 2\pi/N, \cdot\cdot\cdot, 
2(N-1)\pi/N)
,
\label{WHN}
\end{equation}
where the gauge fields $A_2$ is the composite gauge field 
defined in Eq.~(\ref{eq:composite_gauge_field}) of the 
${\mathbb C}P^{N-1}$ model. 

The simplest solution of the BPS equation (\ref{eq:BPSeq}) 
satisfying the $Z_N$ twisted boundary condition in 
Eq.~(\ref{ZNC}) is given by 
\begin{eqnarray}
&& \omega_L
 = \left(0, \cdots, 0,1,\lambda 
e^{+2\pi z/N} , 0,\cdot\cdot\cdot \right)^{T}\,, \quad
 \omega_R
 = \left(0,\cdots, 0,1,\lambda 
e^{-2\pi z/N} ,0,\cdot\cdot\cdot,0\right)^{T}\,.\quad
\label{eq:fractional_instanton}
\end{eqnarray}
One can easily see that these BPS solutions give the total action 
$S$ and the topological charge $Q$ as 
\begin{equation}
S=1/N,\,\,\,\,Q=\pm 1/N\,.
\label{Ini}
\end{equation}
Therefore we call this BPS solution as a fractional instanton. 
The fractional instanton is located at 
$x_1=\frac{N}{2\pi}\log\frac{1}{\lambda}$. 

A neutral bion is a molecule of fractional 
instantons and anti-instatons and is unstable under the 
annihilation process. 
Then we anticipate that generic neutral bion 
configuration is not a solution of field equations. 
Therefore we wish to construct a field configuration that reduces 
to a solution at least for far-separated fractional instantons 
and anti-instantons asymptotically. 

From the BPS solution in Eq.(\ref{eq:fractional_instanton}) 
and their complex conjugates, we are naturally led to consider 
the following ansatz for the ${\mathbb C}P^{1}$ model 
satisfying the ${\mathbb Z}_{N}$ twisted boundary condition 
in Eq.~(\ref{ZNC}) 
\begin{equation}
\omega
 = \left(0,\cdot\cdot\cdot,0,1+\lambda_{2}
e^{2\pi(z+\bar{z})/N},\,\,\lambda_{1}
e^{2\pi z/N},0,\cdot\cdot\cdot, 0 \right)^{T}\,. 
 \label{BBozeroN}
\end{equation}
A fractional instanton is located at 
$x_1=\frac{N}{2\pi}\log\frac{1}{\lambda_1}$, 
and a fractional anti-instanton is at 
$x_1=\frac{N}{2\pi}\log\frac{\lambda_1}{\lambda_2}$, respectively. 
This configuration becomes a solution of field equations asymptotically 
as the separation $\frac{N}{2\pi}\log\frac{\lambda_1^2}{\lambda_2}$ 
goes to infinity. 
For $\lambda_{1}^{2}\gg \lambda_{2}$, this configuration 
corresponds to a $1/N$ instanton ($S=1/N$, $Q=1/N$) 
and a $1/N$ anti-instanton ($S=1/N$, $Q=-1/N$) at large separations. 
The total action and the net topological charge in the large-separation 
limit are given by
\begin{equation}
S=2/N,\,\,\,\,Q=0\,.
\label{Ini}
\end{equation}
The topological charge is unchanged as the separation between 
the fractional instanton and the anti-instanton changes. 
However, the total action monotonically decreases as the 
instantons get closer, which shows an attractive force.

\section{Interactions between fractional instanton and anti-instanton}

We have evaluated analytically the total action for various 
values of separation between the fractional instanton and 
anti-instanton. 

\begin{figure}[htbp]
\begin{center}
 \includegraphics[width=0.99\textwidth]{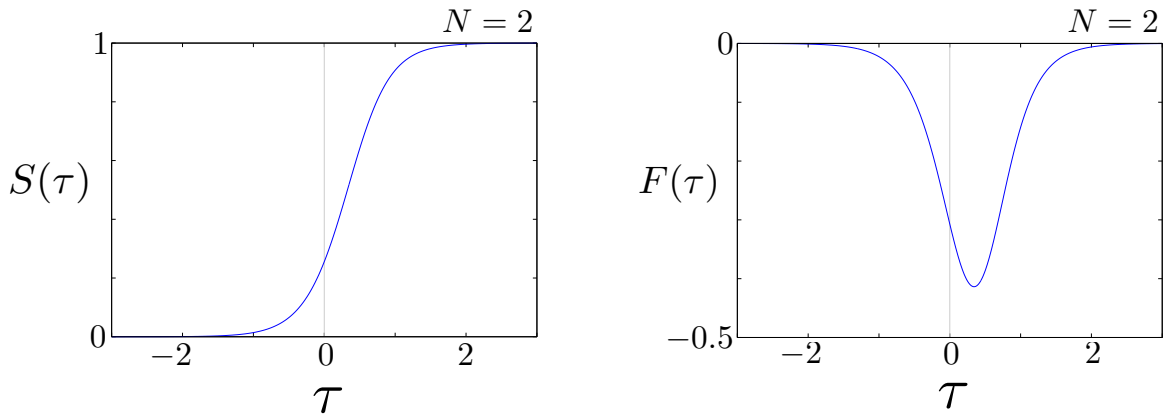}
\end{center}
\caption{The $\tau=(1/\pi)\log\lambda_{1}^{2}/\lambda_{2}$ 
dependence of the total action $S$ 
and the force $F=-{dS\over{d\tau}}$ for 
(\ref{BBozeroN}) with $N=2$. 
For $\tau\geq0$, we can interpret $\tau$ as separation between 
the instanton constituents.
The configuration is changed from $S=1$ to $S=0$, due to the 
attractive force.
The configuration for $\tau\geq 1$ corresponds to neutral bions.}
\label{BBSFlog}
\end{figure}

In Fig.~\ref{BBSFlog}, we show the total action $S$ as a 
function of the separation $\tau=(1/\pi)\log\lambda_{1}^{2}/\lambda_{2}$ 
for $N=2$. 
We also show the attractive force defined by $F=-{dS\over{d\tau}}$ 
as a function of the separation $\tau$. 
For negative values, the moduli parameter $\tau$ loses the 
physical meaning as the separation between the fractional instanton 
and anti-instanton, and merely describes a deformation of the 
field configuration. 

To compare our concrete ansatz (\ref{BBozeroN}) to 
the far-separated instanton argument 
in Ref.~\cite{Dunne:2012ae}, 
we analyze the interaction part of the action for 
our configuration.
The interaction part of the action density is written 
as the difference of the action density $s(x_{1})$ compared to 
the sum of the one fractional-instanton density and one 
fractional-anti-instanton density 
$s_{\nu=1/N}(x_{1})+s_{\nu=-1/N}(x_{1})$,
\begin{equation}
s_{\rm int}(x_{1}) = s(x_{1})-(s_{\nu=1/N}(x_{1})+s_{\nu=-1/N}(x_{1}))\,.
\end{equation}
The total interaction action is given by integrating over the space 
\begin{equation}
S_{\rm int}(N, \tau)\, =\, {1\over{ \pi}}\int dx\, s_{\rm int}(x_{1})\,.
\end{equation}

\begin{figure}[htbp]
\begin{center}
 \includegraphics[width=0.99\textwidth]{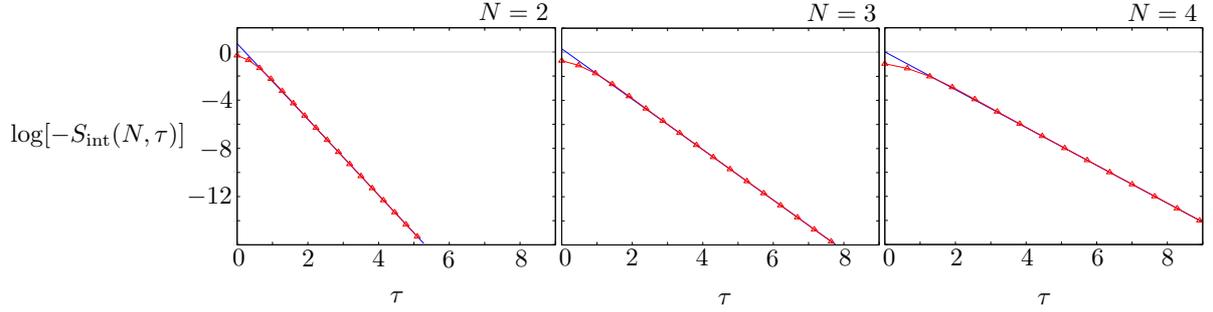}
\end{center}
\caption{Plot of $\log (-S_{\rm int}(N,\tau))$ as a function of $\tau$ 
for $N=2$ (left), $N=3$ (center) and $N=4$ (right) for 
(\ref{BBozeroN}) (red curves with 
triangle points).
For $\tau>1$, the curve is almost equivalent to 
$-(2\pi/N)\tau+ C(N)$ (blue curves).}
\label{Sint}
\end{figure}

\begin{figure}[htbp]
\begin{center}
 \includegraphics[width=0.6\textwidth]{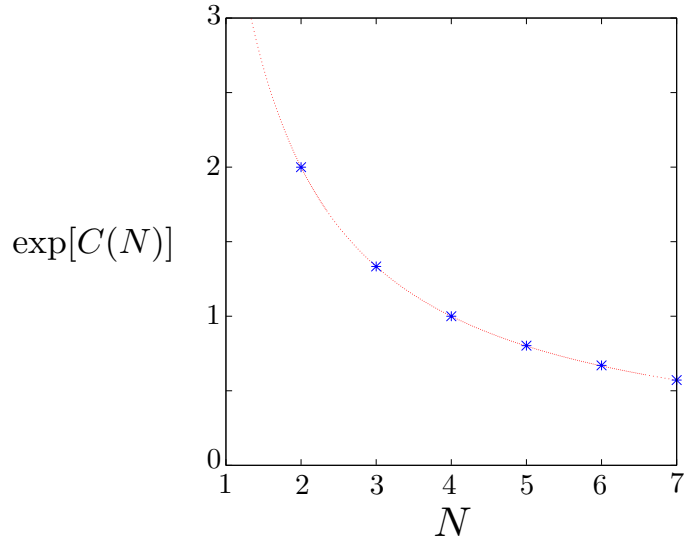}
\end{center}
\caption{The coefficient of the interaction action $\exp[C(N)]$ 
in Eq.(\ref{eq:int_action_approx}) 
as a function of $N$ for $N = 2,3,4,5,6,7$ for the 
Ansatz (\ref{BBozeroN}) (blue points). 
The coefficient can be approximated by $4/N$ (a red curve).}
\label{N-dep}
\end{figure}

In Fig.~\ref{Sint}, we plot the logarithm of the total 
interaction action $S_{\rm int}(N, \tau)$ as a function of $\tau$
for $N=2,3,4$.
In the $\tau\geq1$ region, $\log (-S_{\rm int}(N,\tau))$ can be 
well approximated by analytic lines,
\begin{equation}
\log \left[-S_{\rm int}(N,\tau)\right]\,\,\sim\,\, -\xi(N)\, \tau \,\,+\,\, C(N)\,,\,\,\,\,\,\,\,\,\,\,\,\,\,\,\,\,\, (\tau\geq 1)\,,
\end{equation}
where $\xi(N)$ is a slope and $C(N)$ is a $y$-intercept.
In Fig.~\ref{Sint} we simultaneously depict these analytic lines for the three cases. 
The slopes $\xi$ of the approximate lines 
read $\xi\sim \pi$ for $N=2$, $\xi\sim 2\pi/3$ for $N=3$ 
and $\xi\sim \pi/2$ for $N=4$,
which indicates that the slope $\xi$ can be generally expressed as
\begin{equation}
\xi(N)\,\sim\, {2\pi\over{N}}\,.
\end{equation}
Therefore we observe 
that the interaction action can be written as the following 
form for $\tau\geq1$ region,
\begin{equation}
S_{\rm int} (N, \tau)\, \,\sim\,\, -\,e^{C} \, e^{-\xi \tau}\,, \,\,\,\,\,\,\,\,\,\,\,\xi={2\pi\over{N}}\,,
\,\,\,\,\,\,\,\,\,\,\,\,\,\,\,\,\, (\tau\geq 1)\,.
\label{eq:int_action_approx}
\end{equation}
This $\xi$ is equivalent to the (dimensionless) lowest Kaluza-Klein spectrum $Lm_{LKK}$,
which is given as $Lm_{LKK} = |q_{i}-q_{j}|=2\pi/N$, where $q_{i}$ and $q_{j}$
are two nonzero components of Wilson-loop holonomies in (\ref{WHN}).
Restoring the coupling constant $g^2$, we finally obtain that 
the fractional instanton and anti-instanton exert attractive 
interaction from large to small separations as 
\begin{equation}
S_{\rm int}(\tau)=- \frac{8\xi}{g^{2}}e^{-\xi\tau}\,,
\,\,\,\,\,\,\,\,\,\,\,\,\,
\xi\equiv {2\pi\over{N}}\,.
\label{bionS1}
\end{equation}
This is precisely in accord with the approximate result 
for far-separated case \cite{Dunne:2012ae}. 


\section{Bion amplitudes}

Now we are in a position to sketch briefly how our results fit 
into the resurgence phenomenon.

According to the study, the imaginary ambiguity 
arising in non-Borel-summable perturbative series is 
compensated by the contributions of neutral bions. 
This phenomenon, which is called ``resurgence", works as 
follows \cite{Dunne:2012ae}:
The effective interaction energy by bosonic exchange between 
one fractional instanton and one fractional 
anti-instanton is given in Eq.~(\ref{bionS1}). 
The total bion amplitude including the fermion zero-mode exchange contribution 
is given by
\begin{equation}
\mathcal{B} \propto -e^{-2S_{I}/N}\int_{0}^{\infty} d\tau\, e^{-V_{\rm eff}^{ij}(\tau)}\,,
\end{equation}
with $V_{\rm eff}
(\tau)= S_{\rm int}(\tau)+2N_{f}\xi\tau$ and $S_{I}$ being 
the instanton action. $N_{f}$ stands for fermion flavors.
For neutral bion, 
semiclassical description of independent fractional instantons breaks down 
since the interaction is attractive and instantons are merged 
in the end. 
Here, the BZJ-prescription, replacing $g^{2}\to-g^{2}$, 
works to extract meaningful information from this amplitude.
The prescription turns the interaction (spuriously) into a 
repulsive one and the amplitude becomes well-defined as
\begin{equation}
\mathcal{B}
(g^{2}, N_{f})\,\,\,\to\,\,\,\tilde{\mathcal{B}}
(-g^{2}, N_{f})\,\,\,\propto\,\,\, 
(-g^{2}N/8\pi)^{2N_{f}}\Gamma(2N_{f})e^{-2S_{I}/N}\,.
\end{equation}
By using the analytic continuation in 
the $g^{2}$ complex plane, we can continue back to the original $g^{2}$.
For $N_{f}=0$ case, we then encounter the following imaginary ambiguity in the amplitude as
\begin{equation}
\tilde{\mathcal{B}}
(g^{2}, 0) \,\,\,\propto\,\,\,\left( \log(g^{2}N/8\pi)
-\gamma \pm i\pi\right)
e^{-2S_{I}/N}\,.
\end{equation}
We can rephrase this situation as follows: 
unstable negative modes of bions give 
rise to imaginary ambiguities of the amplitude.
The imaginary ambiguity has the same magnitude with an opposite 
sign as the leading-order ambiguity ($\sim \mp i\pi e^{-2S_{I}/N}$) 
arising from the non-Borel-summable series expanded around 
the perturbative vacuum. 
The ambiguities at higher orders 
($\mp i\pi e^{-4S_{I}/N}$, $\mp i\pi e^{-6S_{I}/N}$,...) 
are cancelled by amplitudes of bion molecules (2-bion, 3-bion,...), 
and the full trans-series expansion around 
the perturbative and non-perturbative vacua results in unambiguous 
definition of field theories.

Finally we would like to mention the results of our more recent study 
of bions in the Grassmann sigma model \cite{Misumi:2014bsa}. 
The Grassmann sigma model with the Grassmann manifold 
$Gr_{N_{\rm F},N_{\rm C}}$ as a target space include 
the ${\mathbb C}P^{N_{\rm F}-1}$ model as a subclass with 
$N_{\rm C}=1$ (and $N_{\rm F}-N_{\rm C}=1$). 
We have found that charged bions do not exist in the 
${\mathbb C}P^{N_{\rm F}-1}$ model, whereas the 
genuine Grassmann sigma model with 
$N_{\rm C} \geq 2, N_{\rm F}-N_{\rm C}\geq 2$ admits charged bions. 
The Grassmann sigma model admits BPS 
fractional instantons \cite{Eto:2006mz,Eto:2007aw,Eto:2006pg} 
with instanton number greater than unity 
(of order $N_{\rm C}$), which cannot be reduced to composite 
of instantons and fractional instantons. 
Interaction between fractional instanton and anti-instanton 
 in the Grassmann model is obtained to form neutral bions. 
We have also obtained exact non-BPS solutions of field equations 
for charged bions. 
To obtain these results, D-brane configurations are found to 
be a valuable tool to analyze fractional instantons and bions.

\ack
We are grateful to Mithat \"{U}nsal and Gerald Dunne for their 
interest and valuable comments and correspondences on their 
related work during the entire course of our study. 
T.\ M.\ and N.\ S.\ thank Philip Argyres, Alexei Cherman, Falk Bruckmann, 
and Tin Sulejmanpasic and other participants of CERN theory 
institute 2014, ``Resurgence and Transseries in quantum, 
gauge and string theories'' for the fruitful discussion 
and useful correspondence. 
T.\ M.\  is in part supported by the Japan Society for the 
Promotion of Science (JSPS) Grants Number 26800147. 
The work of M.\ N.\ is supported in part by Grant-in-Aid for 
Scientific Research (No. 25400268) and by the ``Topological 
Quantum Phenomena''  Grant-in-Aid for Scientific Research on 
Innovative Areas (No. 25103720) from the Ministry of Education, 
Culture, Sports, Science and Technology  (MEXT) of Japan.
N.\ S.\  is supported by Grant-in Aid for Scientific Research 
No. 25400241 from the Ministry of Education, 
Culture, Sports, Science and Technology  (MEXT) of Japan.

\end{document}